\documentclass[aip,apl,reprint]{revtex4-1}
\usepackage{amssymb}
\usepackage{amsmath}
\usepackage{dcolumn}
\usepackage{bm}
\usepackage{graphicx}

\begin{document}

\title{Broadband opto-mechanical phase shifter for photonic integrated circuits}
\author{Xiang Guo}
\affiliation{Key Lab of Quantum Information, University of Science
and Technology of China, Hefei 230026, Anhui, P. R. China}
\author{Chang-Ling Zou}
\email{changlingzou@gmail.com}\affiliation{Key Lab of Quantum Information, University of Science
and Technology of China, Hefei 230026, Anhui, P. R. China}
\author{Xi-Feng Ren}
\email{renxf@ustc.edu.cn} \affiliation{Key Lab of Quantum Information, University of Science
and Technology of China, Hefei 230026, Anhui, P. R. China}
\author{Fang-Wen Sun}
\affiliation{Key Lab of Quantum
Information, University of Science and Technology of China, Hefei
230026, Anhui, P. R. China}
\author{Guang-Can Guo}
\affiliation{Key Lab of Quantum Information, University of Science
and Technology of China, Hefei 230026, Anhui, P. R. China}

\begin{abstract}
A broadband opto-mechanical phase shifter for photonic integrated circuits is proposed and numerically investigated. The structure consists of a mode-carrying waveguide and a deformable non-mode-carrying nanostring, which are parallel with each other. Utilizing the optical gradient force between them, the nanostring can be deflected. Thus the effective refractive indices of the waveguide is changed with the deformation, and further causes a phase shift. The phase shift under different geometry sizes, launched powers and boundary conditions are calculated and the dynamical properties as well as the thermal noise's effect are also discussed. It is demonstrated that a $\pi$ phase shift can be realized with only about 0.64 mW launched power and 50 $\mu m$ long nanostring. The proposed phase shifter may find potential usage in future investigation of photonic integrated circuits.
\end{abstract}

\maketitle

Photonic integrated circuit, composed of multiple photonic devices in a
single chip, has attracted more and more attention due to its small
footprint, reduced power consumption and enhanced processing stability.
It has shown potential usages in both ultrafast classical\cite{Lipson05,Lipson08} and quantum information processing\cite{Brien,Ladd}. Special designs is needed for optical components that can be minimized in photonic integrated circuit, such as the surface plasmon based polarizer\cite{Dong} and polarization beam splitter\cite{Zou}. With recent advances in nano-fabrication technology, opto-mechanics utilizing optical forces are of great interests for novel devices in photonic integrated circuits. The optical gradient forces between two coupled waveguides have been illustrated theoretically and experimentally in many works\cite{Povinelli,Pernice,Zhu,Pernice2,Rakich,Ma,M.Li,Roels}.

Among the different optical components, phase shifter is an important issue in the field of integrated photonics. In traditional optical circuits, phase shift can be easily achieved by modulating the distance between the optical elements. However,  in the integrated photonics, components are fixed and connected by the waveguide, whose length is hard to change. One way to control the phase is changing the effective refractive index of the waveguide, which has been realized by electron carrier injection\cite{Tsang,Almeida,Xu} and thermal-optical effect\cite{Cocorullo,Pruessner,Song}. However, there are some disadvantages for these methods, such as not relatively big geometry sizes and long relaxation time. More recently, Fong \emph{et al.}\cite{Fong} suggested a coupled waveguide structure in which the optical gradient force is used to deform the waveguides and thus tune their refractive indices. Because of the coupling of the waveguide, light will transfer from one waveguide to the other along the propagation process, which makes the device relatively complex. Further more, it is usually not easy to selectively excite the desired mode in the waveguides, which makes the force between the two waveguides vary from attractive to repulsive according to phase difference between the incoming lightwaves\cite{Pernice2}.

Here, we propose a all-optical phase shifter by opto-mechanical force in photonic integrated circuits. A non-mode-carrying nanostring is used to make sure that the light will always transport along the mode-carrying waveguide and optical forces will be attractive all the time. It is demonstrated that large phase shift can be achieved in small geometry size and low launched power using this all-optical structure. We set different boundary conditions for our structure and find one of them can further decrease the pump power. An important character of our design is the broadband property, which means our structure is effective for a broadband signal light. Finally, the phase shifter's dynamical properties and thermal noise's effects are discussed.

\begin{figure}[tbp]
\centerline{
\includegraphics[width=0.36\paperwidth]{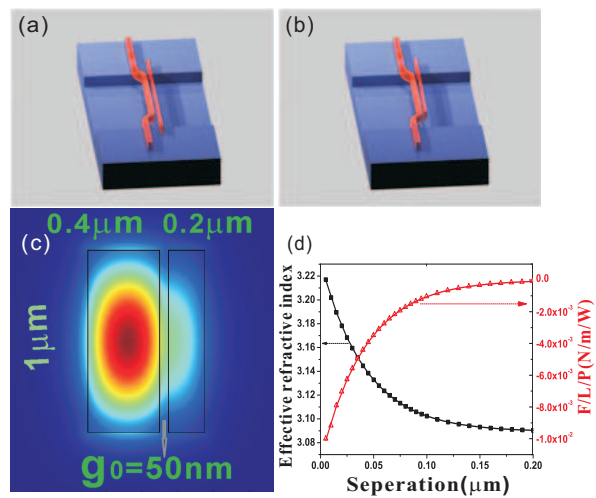}}
\caption{(color online) Illustration of the designed phase shifter. The mode-carrying waveguide is paralleled by a non-mode-carrying nanostring. (a) The nanostring is double-clamped. (b) The nanostring is set to be a cantilever. (c) Mode profile in cross-section. (d) The effective refractive index of the waveguide and the optical force in dependence of the gap.}
\end{figure}

As shown in Fig. 1, a deformable nanostring is put in parallel with the 'controlling' part of a waveguide. Fig. 1a and 1b shows two possible geometries, where the deformable nanostring can be a double-clamped beam or nanocantilever. At the head and tail of this part, the waveguide has curved shape, thus the waveguide can approach the nanostring adiabatically and avoid the light scattering by nanostring end faces \cite{Zou2}. The cross section of the phase shifter is shown in Fig. 1c. The height and width of waveguide are 1$\mu m$ and 0.4$\mu m$, while for the nanostring they are 1$\mu m$ and 0.2$\mu m$ respectively. The width of nanostring is much thinner, thus the nanostring is easier to be deflected, and the light can not leak out through the nanostring. When a pump light is launched into the phase shifter, the optical gradient force will be generated, which can attract and deform the nanostring.  Thus the effective refractive index of the waveguide varied according to the distance between waveguide and nanostring. A signal light transferring through the same waveguide will acquire a phase shift, which can be adjusted by tuning the power of pump light. Obviously, the power of the signal light must be low enough to avoid generating an additional force. This condition can be fulfilled when we use this phase shift in quantum photonic chips\cite{Brien,Ladd}, where the intensity of the signal light is on single photon level. Unlike those structures that take advantage of the cavity resonance, which have strong restrictions in the operating wavelength, our designed phase shifter can be applied to broadband signal light as the effective refractive index and optical force are not sensitive to light's wavelength. In the following studies, the pump light's wavelength is $\lambda _{p}=1550nm$, the signal light's wavelength is $\lambda_{s}=1480nm$, and the initial gap between waveguide and the nanostring is $g_{0}=50nm$ for the sake of simplification. The device is made up on a silicon chip.

The gradient optical force near a waveguide can be describe as by\cite{Rakich,Ma}:
\begin{equation}
\frac{F}{LP}=\frac{1}{C}\frac{\partial n_{eff}}{\partial d},
\end{equation}
where $F$ stands for the optical forces, $L$ for the length of the nanostring and $P$ for the launched power of light respectively. The effective refractive index $n_{eff}$ in dependence to the gap $d$ can be numerically solved by Finite Element Method(COMSOL Multiphysics), and then we can calculate the optical force according to Eq. (1). The result can be seen in Fig. 1d, and these calculation data are fitted by exponential curve for the convenience of following analysis.

Now, we can calculate the waveguide and nanostring's deflection due to the optical forces. According to the Euler-Bernoulli beam theory\cite{Cleland}, deflection is determined by the equation:
\begin{equation}
\frac{d^{4}u}{dz^{4}}=\frac{12}{Et^{2}}f(z)=\frac{12}{Et^{2}}\cdot
\frac{F}{LP}\cdot \frac{P}{A},
\end{equation}
where $E=131\times 10^{9}Pa$ stands for the silicon's Young's modulus, $A=0.2\mu m^{2}(0.4\mu m^{2}$) for the nanostring's(waveguide's) cross-section area and $t=0.2\mu m(0.4\mu m$) for nanostring's(waveguide's) width and f(z) is the volume density of force. There are two possible boundary conditions for the nanostring: (1) Double-clamped beam, both the waveguide and the nanostring obey the same boundary conditions: $u(0)=\frac{du}{dz}(0)=u(l)=\frac{du}{dz}(l)=0$. (2) Cantilever, Boundary conditions for the nanostring now are: $u(0)=\frac{du}{dz}(0)=\frac{d^{2}u}{dz^{2}}(l)=\frac{d^{3}u}{dz^{3}}(l)=0$\cite{Cleland}. Combine the equations for optical force and deflection, we can obtain the analytical solutions under these boundary conditions.

\begin{figure}[tbp]
\centerline{
\includegraphics[width=0.37\paperwidth]{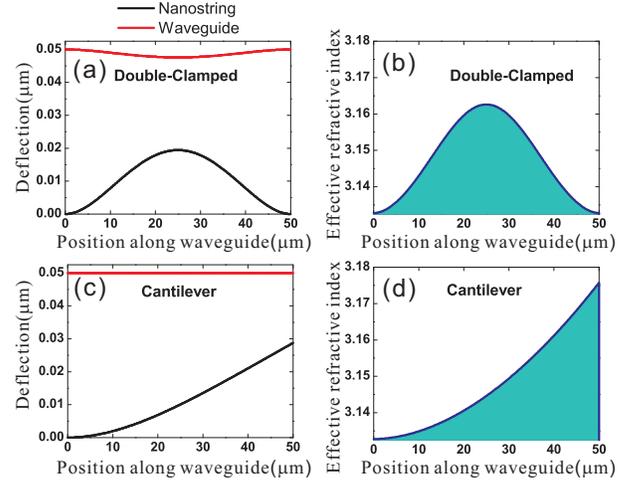}}
\caption{(color online) (a) Deflection curve of the phase-shifter in the case
of double-clamped beam($l=50\mu m, P=21.4mW$). (b) The effective refractive index along the waveguide in the case of double-clamped beam($l=50\mu m, P=21.4mW$). The light-blue shaded area corresponds to the delta optical length in this situation. (c) Deflection curve of the phase-shifter in the case of cantilever($l=50\mu m, P=0.64mW$). (d) The effective refractive index along the waveguide in the case of cantilever($l=50\mu m, P=0.64mW$). The light-blue shaded area corresponds to the delta optical length in this situation.}
\end{figure}

Fig. 2a and 2b show the results for double-clamped beam with $l=50\mu m,P=21.4mW$. Fig. 2a shows the deflection curve of the nanostring and the waveguide. As the waveguide is thicker than the nanostring, it has a smaller deflection. We can see that the maximum deflection happens in the middle of both beams with a value of 19.7nm for the nanostring and 2.4nm for the waveguide. Fig. 2b shows the effective refractive index along the waveguide under this condition. As
is shown clearly that the effective refractive index along the waveguide
will vary according to the deformation and hence different gap values
between the waveguide and the nanostring. With this relationship, we can finally obtain the changed optical length with the equation $\Delta L=L^{\prime }-L=\int n_{eff}(z)dz-n_{eff}(g_{0})\times l$. This value is shown in Fig. 2b by the area of the light-blue shaded region, corresponding to an optical length shift of $\Delta L=0.7418\mu m$. As the signal light's wavelength is 1480nm, the phase shift $\varphi=\Delta L/\lambda\times 360^{\circ }=180.4^{\circ }$ for the signal light. For the nanocantilever, the deflection curve and effective refractive index along the waveguide are shown in Fig. 2(c) and Fig. 2(d), with $l=50\mu m,P=0.64mW$. We can find in Fig. 2c that the max deflection happens in the free end of the cantilever with a value of 28.8nm, less than the initial gap of 50nm. We get the optical length shift in this situation to be $\Delta L=0.7507\mu m$ and $\varphi =182.6^{\circ }$ for signal light.

Now we perform a detailed study of the phase shifter. Fig. 3a shows the dependence of phase shift to the power when the nanostring's length is fixed to $l=50\mu m$. It is obviously that cantilever can perform a phase shift with much smaller launched power than the double-clamped one.  For example, in order to get a phase shift of 180$^{\circ}$, we need a small pump light power of only about 0.64mW in the case of cantilever, while 21.4mW for the double-clamped structure. In addition, with increasing pump power, we can continuously control the signal light phase shift from 0 to 180$^{\circ}$. Fig. 3b shows the dependence of phase shift to the nanostring's length when the launched power is fixed to $
P=0.64mW$. It shows that the value of phase shift strongly depends on the nanostring's length. Furthermore, we make contour figures to show the phase shift in dependence of length and power in both boundary conditions in Fig. 3c and Fig. 3d. These results imply that the value of phase shift will be more sensitive to the power and the nanostring's length when phase shift is bigger. Thus, the cascade combination of the phase shifters is much better to form a bigger one. Each nanostring helps produce a smaller phase shift, which is relatively stable, to form a bigger phase shift by the total phase shifter. On the other hand, it can also decrease the geometry size along the nanostring's longitudinal direction and make the structure to be more space-efficient, which is important for integrated circuit.

\begin{figure}[tbp]
\centerline{
\includegraphics[width=0.38\paperwidth]{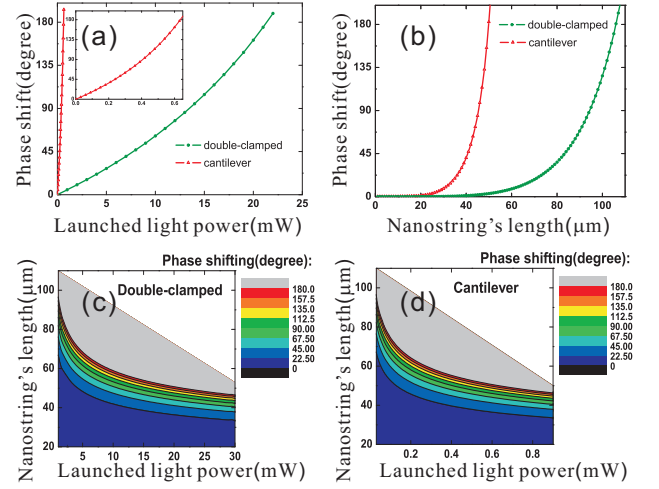}}
\caption{(color online) (a) Dependence of phase shift to the launched power
under the two different boundary conditions(while nanostring's length is fixed to 50$
\protect\mu$m). (b) Dependence of phase shift to the nanostring's length under
the two different boundary conditions (while launched power is fixed to
0.64mW). (c) Dependence of phase shift to the launched power and nanostring's
length in the case of double-clamped beam. (d) Dependence of phase shift
to the launched power and nanostring's length in the case of cantilever.}
\end{figure}

In the information process, the response time of the phase shift device is very important. Actually, when we suddenly launch the pump light to stretch the nanostring, it can not reach the balanced position immediately. It will vibrate and the vibration will decay for a period of time. According to the Euler-Bernoulli beam theory\cite{Cleland}, we get the formula that $\omega _{n}=\sqrt{\frac{EIy}{\rho A}}\beta _{n}^{2},$ where $\beta _{n}$ has a certain series of values according to different boundary conditions, which will be shown later. For beam has a limited lifetime, the vibration frequency would be a complex number: $\omega _{n}^{\prime }=(1+\frac{i}{2Q})\omega _{n}$, where Q is the quality factor. The imaginary part stands for the decay of the n$_{th}$ mode of the beam, whose amplitude will decay with $e^{-\omega _{n}t/2Q}$. So the lifetime $\tau $ should be:
\begin{equation}
\begin{split}
\tau =\frac{2Q}{\omega _{n}}=\frac{2Q}{\sqrt{\frac{EIy}{\rho
A}}\beta _{n}^{2}}=\frac{2Ql^{2}}{\sqrt{\frac{Et^{2}}{12\rho }}(\beta
_{n}l)^{2}},
\end{split}
\end{equation}
where $\rho=2330kg/m^{3}$ stands for the mass density of the nanostring. In the case of $l=50\mu m$, $\tau =$ $\frac{Q}{(\beta_{n}l)^{2}}\times 3.6524\mu s$. For double-clamped one, $\beta _{n}l$ obeys the equation that $\cos (\beta _{n}l)\cosh (\beta _{n}l)-1=0$. The solutions to the equation are $\beta _{n}l=$0, 4.73004, 7.8532, 10.9956,
14.1372... For the fundamental vibration mode, $\beta _{1}l=4.73004$. As can be seen from Eq. (3), the decaying constant will decrease with the mode's order. So the first mode will make the most contribution to the vibration. For $Q=100$, the lifetime for double-clamped beam is $\tau_{double-clamped}=16.3\mu s$. For the cantilever, the equation is $\cos (\beta _{n}l)\cosh (\beta _{n}l)+1=0$. The solutions are $\beta _{n}l=$1.875, 4.694, 7.855, 10.996... $\beta_{1}=1.875$ for fundamental mode, corresponding $\tau _{cantilever}=103.9\mu s$, which is more than 6 times bigger than $\tau _{double-clamped}$. This means the response of cantilever will be slower than double-clamped beam.

Finally, we will discuss the thermal noise limited performance of the opto-mechancial phase shifter. From thermal dynamics, single degree of freedom corresponds to an energy of $\frac{1}{2}k_{B}T$, so approximately $\frac{1}{2}ku_{\max }^{2}=E_{k}+E_{p}=k_{B}T$, where $u_{max}$ is the amplitude of thermal motion of the nanostring. The elastic constant $k$ can be derived by $F=\frac{F}{LP}*P*l=ku$, where $u$ stands for the average deflection of the beam. By calculating the $u$ against $P$, we linearly fitted the slopes and obtained $k_{double-clamped}=0.4971N/m,k_{cantilever}=0.0133N/m$ with $l=50\mu m$. Then we can work out the vibration amplitude caused by the thermal noise are (when temperature T=300K): $u_{double-clamped}=0.1291nm,u_{cantilever}=0.7903nm$, which will cause phase shift uncertainty of about 1.6$^{\circ }$ and 10.1$^{\circ }$ for each. For future practical application of photonic integrated circuits, it may be placed in a low temperature environment as many devices will have a better performance, such as the superconducting detector and polarization beam splitter \cite{zou}. When the temperature is cooled down to 10K, the uncertainty will be decreased to about 0.3$^{\circ }$ and 1.8$^{\circ }$ respectively. The results show that the cantilever is more sensitive to the thermal noise than the double-clamped beam.

In conclusion, we have designed an all-optical phase shifter based on opto-mechanical structure for photonic integrated circuits. The phase shift results under different launched powers and geometry sizes were discussed. It was found that the device can control the light phase with low power, small geometry size and quick relaxation time, and a cascade of these shifters can generate even better results. The thermal-noise's effects were also discussed to meet the real environment. Our result may have a potential usage in the future classical and quantum photonic integrated circuits.

We are grateful to Xiang-Dong Chen for his help in preparing Fig. 1. This work was funded by the National Basic Research Program of China (Grants
No. 2011CBA00200 and 2011CB921200), the Innovation funds from Chinese
Academy of Sciences (Grants No. 60921091), the National Natural Science
Foundation of China (Grants No. 10904137), the
Fundamental Research Funds for the Central Universities (Grants No.
WK2470000005) and the Program for New Century Excellent Talents in University.

\end{document}